\documentclass[manuscript]{aastex62}
\usepackage{mhchem}
\usepackage[utf8]{inputenc}
\usepackage[T1]{fontenc} 
\usepackage{comment}

\graphicspath{{./}{figures/}}

\received{January 1, 2018}
\revised{January 7, 2018}
\accepted{\today}
\submitjournal{ApJ}

\shorttitle{\ce{H2C3O} Isomers, Revisited}
\shortauthors{Shingledecker et al.}

\newcommand{\new}[1]{\textcolor{black}{#1}}

\begin{document}

\title{The Case of \ce{H2C3O} Isomers, Revisited: Solving the Mystery of the Missing Propadienone}

\correspondingauthor{Christopher N. Shingledecker}
\email{cns@mpe.mpg.de}

\author[0000-0002-5171-7568]{Christopher N. Shingledecker}
\affil{Center for Astrochemical Studies \\
Max Planck Intitute for Extraterrestrial Physics \\
Garching, Germany}
\affil{Institute for Theoretical Chemistry \\
University of Stuttgart \\
Pfaffenwaldring 55, 70569 \\
Stuttgart, Germany}

\author[0000-0002-0175-0845]{Sonia~\'Alvarez-Barcia}
\affil{Institute for Theoretical Chemistry \\
University of Stuttgart \\
Pfaffenwaldring 55, 70569 \\
Stuttgart, Germany}

\author{Viktoria H. Korn}
\affil{Institute for Theoretical Chemistry \\
University of Stuttgart \\
Pfaffenwaldring 55, 70569 \\
Stuttgart, Germany}

\author[0000-0001-6178-7669]{Johannes Kästner}
\affil{Institute for Theoretical Chemistry \\
University of Stuttgart \\
Pfaffenwaldring 55, 70569 \\
Stuttgart, Germany}



\begin{abstract}

To date, two isomers of \ce{H2C3O} have been detected, namely, propynal (HCCCHO) and cylclopropenone (\ce{c-H2C3O}). A third, propadienone (\ce{CH2CCO}), has thus far eluded observers despite the fact that it is the lowest in energy of the three. This previously noted result is in contradiction of the minimum energy principle, which posits that the abundances of isomers in interstellar environments can be predicted based on their relative stabilities - and suggests, rather, the importance of kinetic over thermodynamic effects in explaining the role of such species. 

Here, we report results of \textit{ab initio} quantum chemical calculations of the reaction between H and (a) \ce{HC3O}, (b) \ce{H2C3O} (both propynal and propadienone), and (c) \ce{CH2CHCO}. We have found that, among all possible reactions between atomic hydrogen and either propadienone or propynal, only the destruction of propadienone is barrierless and exothermic. That this destruction pathway is indeed behind the non-detection of \ce{CH2CCO} is further suggested by our finding that the product of this process, the radical \ce{CH2CHCO}, can subsequently react barrierlessly with H to form propenal (\ce{CH2CHCHO}) which has, in fact, been detected in regions where the other two \ce{H2C3O} isomers are observed. Thus, these results not only shed light on a previously unresolved astrochemical mystery, but also further highlight the importance of kinetics in understanding the abundances of interstellar molecules. 

\end{abstract}

\keywords{astrochemistry --- ISM: abundances --- ISM: clouds --- ISM: molecules --- molecular processes }


\section{Introduction} \label{sec:intro}

In a recent survey by \citet{mcguire_census_2018}, it was noted that, to date, approximately 200 different individual molecular species have been detected in either interstellar or circumstellar regions. As observing facilities and observational techniques have become more sophisticated, larger and more complex molecules have been detected, including a number of isomers - one of the most notable being the branched form of propyl cyanide (\ce{C3H7CN}) \citep{belloche_detection_2014}.

Uncovering the chemical mechanisms behind the observed relative abundances of these isomers remains one of the major challenges in astrochemistry today, and as increasingly complex molecules are detected this problem will only grow more acute. For example, the abundances of \ce{H2C3O} isomers have posed a longstanding astrochemical conundrum. The first such species to be detected was propynal (\ce{HCCCHO}), observed by \citet{irvine_identification_1988} in the cold core TMC-1. The cyclic molecule cyclopropenone (\ce{c-H2C3O}) was later seen by \citet{hollis_cyclopropenone_2006} towards Sgr B2(N) using the GBT. There is also a third form, propadienone (\ce{CH2CCO}), which was calculated to be the most stable of the three \citep{kom81,mac95,eke96,sco00,karton_pinning_2014}, however, this last form has thus far proven elusive.

One such attempt to detect propadienone was made by \citet{loomis_investigating_2015}, who searched for isomers of \ce{H2C3O} in archival data from the PRIMOS survey of Sgr B2(N).\footnote{https://www.cv.nrao.edu/PRIMOS} There, they reported detections of cyclopropenone  and propynal  but a non-detection of \ce{CH2CCO}. In that work, Loomis and coworkers interpreted their results as highlighting the influence of some as yet unknown reaction that favored either the production of propynal and cyclopropenone and/or the destruction of propadienone.

More recently, \citet{loison_interstellar_2016} reported the non-detections of propadienone towards a number of starless cores and molecular clouds, despite having detected both cyclopropenone and propynal. Based on these observational results, as well as those from detailed chemical modeling, they too concluded that the abundances of the \ce{H2C3O} molecules were kinetically, rather than thermodynamically controlled. Loison and coworkers \new{speculated that the primary contributing factor behind the observed abundances of these isomers were differing formation routes}. However, the results of their simulations regarding propadienone were mixed, with its calculated \new{abundance} being within the observational error of their upper limit for at least sources with typical densities of $\sim 10^4$ \new{cm$^{-3}$} at relevant timescales. Thus, despite evidence suggesting the impact of some key chemical reactions on the abundance of propadienone, their identities remained mysterious.

Here, we report the results of \textit{ab initio} quantum chemical calculations of the reaction \ce{H + HC3O}, as well as hydrogen additions to a number of related species, including propynal and propadienone. Though this study was motivated by the recent detections of \ce{HC5O} \citep{mcguire_detection_2017} and \ce{HC7O} \citep{mcguire_detection_2017,cordiner_deep_2017}, in performing these calculations we have uncovered the likely chemical pathways responsible for the puzzling lack of \ce{CH2CCO}. 

Secondarily, our findings shed additional light on the validity of using thermodynamics to predict molecular abundances in the ISM, more generally, and highlight the central importance of chemical activation energy barriers, i.e. kinetics. Such an attempt to make sense of observational results was made by Lattelais \textit{et al.}, who proposed a general rule based on a comparison between observational data and the results of quantum chemical calculations  \citep{lattelais_interstellar_2009,lattelais_new_2010,lattelais_differential_2011}. This guideline, which they called the ``minimum energy principle'' (hereafter MEP), \new{states that the relative abundances of different isomers in the same region could be estimated based on the energy differences between them, with the most stable being the most abundant.} However, even in their first work on the MEP, \citet{lattelais_interstellar_2009} noted exceptions, for instance, species with the formula \ce{C2H4O2}. Thus, though the aim of this work is neither to further challenge the already dubious MEP nor to again demonstrate - as we have previously done in \citet{loomis_investigating_2015} - that the \ce{H2C3O} isomers violate said principle, our results do provide further evidence for the centrality of kinetics in understanding the theoretical basis behind interstellar isomer abundances.

The rest of this letter is as follows: in \S \ref{sec:comp} we give an overview of the methods and tools used in this study, in \S \ref{sec:results} our results are presented, and their astrochemical implications are discussed in \S \ref{sec:astrochem}, finally, our conclusions are summarized in \S \ref{sec:conc}. 

\section{Computational Details} \label{sec:comp}

We used unrestricted density functional theory (DFT) with the PW6B95 functional \citep{pw6}
and the def2-TZVP basis set \citep{tzvp}. These calculations were done in
Turbomole \citep{TURBOMOLE} accessed via ChemShell \citep{she03,met14}. The
geometry optimizations and instanton calculations were done in DL-Find
\citep{dlf} through ChemShell. All molecular degrees of freedom were optimized in all cases. No symmetry was imposed on the molecules. For radical-radical reactions, like \ce{H + HC3O}, an unrestricted broken-symmetry wave function was used with overall as many spin-up as spin-down electrons, but finite spin density on both radicals. Although a gas-phase model was used, rate constants were calculated with an implicit surface model \citep{meisner_atom_2017}.  Benchmarks were performed for individual geometries
on the coupled-cluster level UCCSD(T)-F12a \citep{ccf12} with a restricted Hartree--Fock reference and the cc-pVTZ-f12 basis set \citep{pet08} in Molpro \citep{molpro2010}.

Minima and transition state structures were verified by frequency
calculations from numerical Hessians in DL-Find. Energies are reported including the harmonic vibrational zero
point energy. Barrierless processes were identified by starting energy
minimizations from the separated reactants. An optimization ending up
in the product minimum demonstrates a barrierless path. 

Bimolecular rate constants were calculated using instanton theory
\citep{mil75,col77} as implemented in DL-Find \citep{rom11} below the
crossover temperature,

\begin{equation}
    T_\mathrm{c} = \frac{\hbar\omega_\mathrm{b}}{2\pi k_\mathrm{b}},
\end{equation}

\noindent
with $\hbar$ and $k_\mathrm{b}$ being the reduced Planck and Boltzmann constants, 
respectively, and $\omega_\mathrm{b}$ the absolute value of the imaginary frequency 
at the transition state. 
The instanton path was discretized with 40 replicas,
except for the calculations at 55~K and 50~K for the reaction \ce{H + CHCCHO}
$\rightarrow$ \ce{HC3O + H2}, where 78 replicas were used. Convergence with
respect to the number of replicas was confirmed by using more replicas at one
low temperature. Through the use of both a well-established correction \citep{kry13,mcc17a} close to
the crossover temperature and the use of reduced instanton theory
\citep{kry13} above that temperature, a continuous curve over the
full temperature range was achieved.

\section{Results and Discussion} \label{sec:results}

The results of our geometry optimizations of the molecule \ce{HC3O} show that it is not linear. Rather, in agreement
with QCISD calculations by \citet{radref} and \citet{coo95}, we found two main configurations, namely, one with a bend
on the CO-end, and another with the bend on the CH-end, with the latter being more stable. 

\begin{figure}[htbp]
\centering
\includegraphics[width=8cm]{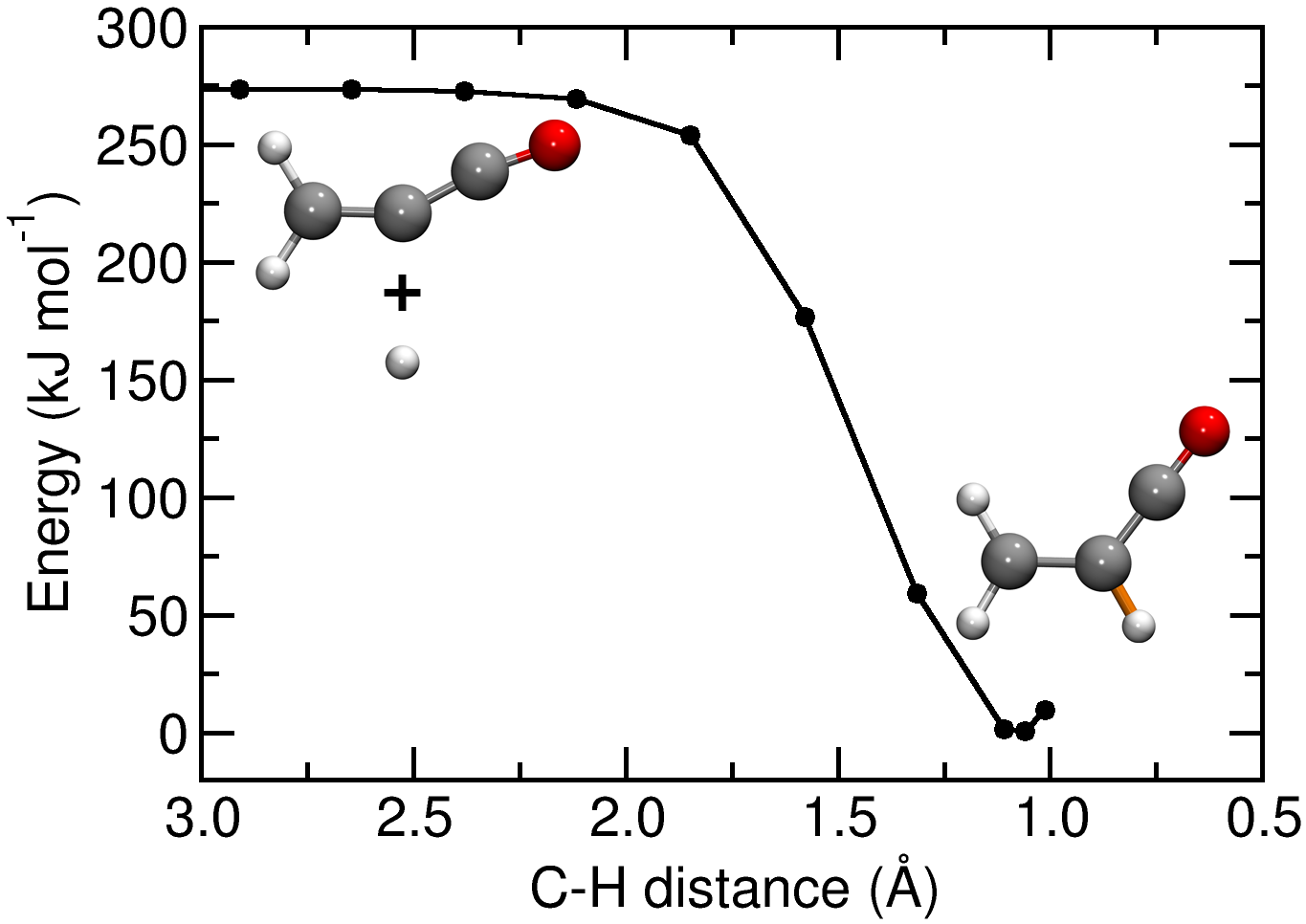}
\caption{Energy along the path of the barrierless reaction (R4), \ce{H + CH2CCO} $\rightarrow$ \ce{CH2CHCO}. The horizontal axis shows the C-H bond length, which is drawn orange in the right structure. Here potential energies without zero point energy are shown.\label{fig:scan}}
\end{figure}

As shown in Table 1, in which the results of our calculations are listed for reactions labeled (R1)-(R13), we further found the first two, 

\begin{equation}
    \ce{H + HC3O -> CH2CCO}
    \label{r1}
    \tag{R1}
\end{equation}

\begin{equation}
    \ce{H + HC3O -> HCCCHO}
    \label{r2}
    \tag{R2}
\end{equation}

\noindent
to be barrierless and exothermic in both the product channels. Thus, both \eqref{r1} and \eqref{r2} should be efficient formation routes for propynal and propadienone on interstellar dust-grain surfaces.
 Using DFT
(PW6B95/def2-TZVP), we found the formation of \ce{CH2CCO} to be exothermic by
363.2~kJ~mol$^{-1}$ and the formation of HCCCHO to be exothermic by
340.3~kJ~mol$^{-1}$. However, the branching ratio between the two is likely to be more sensitive to the orientation of the
reactants than on the exothermicities of these two reactions.
In addition, while \ce{CH2CCO} is more stable than HCCCHO by 22.9~kJ~mol$^{-1}$ using DFT, it is \new{slightly} less stable (by 0.7~kJ~mol$^{-1}$) using CCSD(T)-F12a/cc-pVTZ-F12.
By comparison, \ce{c-H2C3O} is found to be 24.7~kJ~mol$^{-1}$ less stable than \ce{CH2CCO} by DFT and
26.3~kJ~mol$^{-1}$ by CCSD(T)-F12a/cc-pVTZ-F12.
These data are consistent with previous data from W2-F12 theory by \citet{karton_pinning_2014}, who find \ce{CH2CCO} to be the most stable isomer, followed by \ce{HCCCHO} at 2.5~kJ~mol$^{-1}$ and \ce{c-H2C3O} at 29.2~kJ~mol$^{-1}$ and broadly consistent with older calculations at lower levels of theory \citep{kom81,mac95,eke96,sco00}. We find \ce{CH2CCO} to be bent at the $\beta$ carbon atom (see Fig.~\ref{fig:scan}) by 150 degrees, which is close to the 140 degrees found by \citet{sco00} using MP2 theory.

\begin{table}[ht]
  \caption{Overview of reaction energies ($\Delta_\text{r} E$) and activation
    barriers ($\Delta E_{\text{A}}$) in kJ~mol$^{-1}$ of all reactions studied.\label{tab:all}}
\begin{center}
\begin{tabular}{llrr}
\hline
\hline
Label & Reaction & $\Delta_\text{r} E$ & $\Delta E_\text{A}$ \\
\hline
(R1) & \ce{H + HC3O} $\rightarrow$ \ce{CH2CCO} & $-$363.2 & 0.0 \\
(R2) & \ce{H + HC3O} $\rightarrow$ \ce{HCCCHO} & $-$340.3 & 0.0 \\
\hline
(R3) & \ce{H + CH2CCO} $\rightarrow$ \ce{HC3O + H2} & $-$58.2 & 18.4 \\
(R4) & \ce{H + CH2CCO} $\rightarrow$ \ce{CH2CHCO} & $-$246.5 & 0.0 \\
(R5) & \ce{H + CH2CCO} $\rightarrow$ \ce{CH2CCOH} & $-$117.2 & 38.1 \\
(R6) & \ce{H + CH2CCO} $\rightarrow$ \ce{CH2CCHO} & $-$169.7 & $\gg$ \\
(R7) & \ce{H + CH2CCO} $\rightarrow$ \ce{CH3CCO}  & $-$205.4 & 6.2 \\
\hline
(R8) & \ce{H + HCCCHO} $\rightarrow$ \ce{CCCHO + H2} & 128.1 & 128.1\\
(R9) & \ce{H + HCCCHO} $\rightarrow$ \ce{HC3O + H2} & $-$76.5 & 11.7 \\
(R10) & \ce{H + HCCCHO} $\rightarrow$ \ce{CHCCHOH} & $-$168.8 & 26.9 \\
(R11) & \ce{H + HCCCHO} $\rightarrow$ \ce{CHCCH2O} & $-$74.6 & 23.8 \\
(R12) & \ce{H + HCCCHO} $\rightarrow$ \ce{CHCHCHO} & $-$164.5 & 18.9 \\
(R13) & \ce{H + HCCCHO} $\rightarrow$ \ce{CH2CCHO} & $-$192.7 & 11.3 \\

\hline
\hline
\end{tabular}
\end{center}
\end{table}

Thus, while both \ce{CH2CCO} and HCCCHO can be formed by hydrogenation of \ce{HC3O},
they show different stabilities with respect to further hydrogenation. Here, 
we have further studied the hydrogen addition reactions to all atoms of both 
propynal and propadienone, as well as the hydrogen abstraction reaction

\begin{equation}
    \ce{H + CH2CCO -> HC3O + H2}.
    \label{r3}
    \tag{R3}
\end{equation}

\noindent
An examination of the energetics of these processes, summarized in Table~\ref{tab:all},
shows that they are all exothermic. Surprisingly, though, \textit{only} the reaction 

\begin{equation}
    \ce{H + CH2CCO -> CH2CHCO}
    \label{r4}
    \tag{R4}
\end{equation}

\noindent
was found to be barrierless, with  the energy along the path of 
H approaching \ce{CH2CCO} depicted in Fig.~\ref{fig:scan}. 
By contrast, the other channels have barriers larger than
6.2~kJ~mol$^{-1}$. Thus, the formation of \ce{CH2CHCO} is significantly faster
than all alternative channels. Our results also confirm that subsequent reaction between
this radical and atomic hydrogen

\begin{equation}
    \ce{H + CH2CHCO -> CH2CHCHO}
    \label{r5}
    \tag{R14}
\end{equation}

\noindent
is likewise barrierless and results in the formation of propenal, also known as acrolein. 
For reaction (R6), i.e. \ce{H + CH2CCO} $\rightarrow$
\ce{CH2CCHO}, we failed to find an accurate transition state, but the
calculations of the reaction path clearly indicate that the barrier is quite
large.

\begin{figure}[htbp]
\centering
\includegraphics[width=8cm]{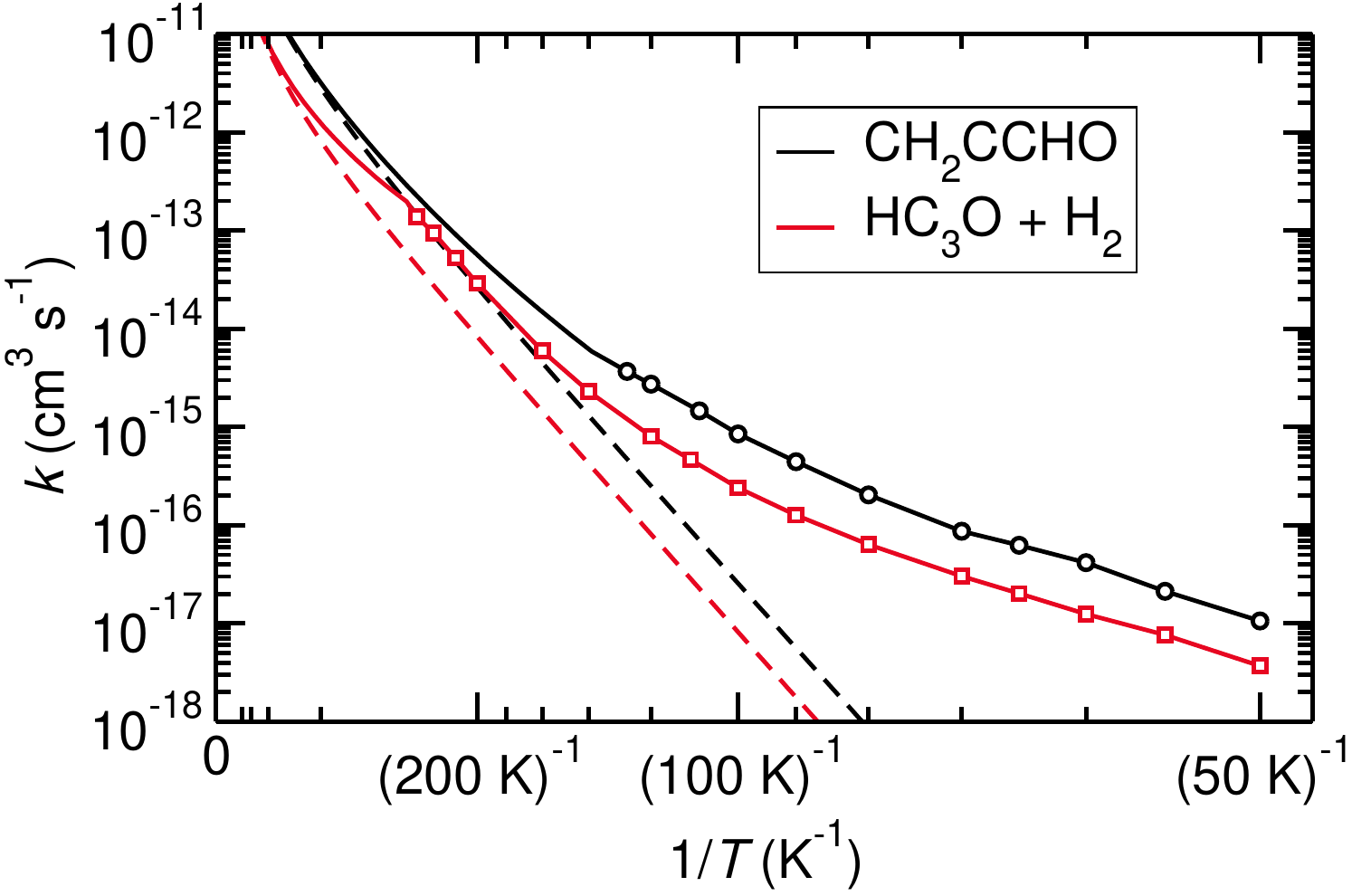}
\caption{Rate constants for the reactions (R9), \ce{H + HCCCHO} $\rightarrow$ \ce{HC3O + H2}
  (red), and (R13), \ce{H + HCCCHO} $\rightarrow$
  \ce{CH2CCHO} (black). The solid lines refer to the instanton rate constants including
  tunneling, the dashed lines ignore tunneling. The open symbols represent the
  individual instanton calculations.\label{fig:rates}}
\end{figure}

As with propadienone, hydrogen atoms can react with propynal, HCCCHO, leading to hydrogen additions
and abstractions. However, unlike the case with \ce{CH2CCO}, 
the results of our calculations show that all of these reactions involving \ce{HCCCHO} 
either have a barrier or (abstraction
of the aliphatic H) are endothermic. Thus, at the low temperatures of
molecular clouds, they will be significantly slower than the barrierless process
destroying \ce{CH2CCO}. Of the six possible reactions, hydrogen abstraction
leading back to \ce{HC3O} (R9) and hydrogen addition leading to \ce{CH2CCHO} (R13) have
the lowest barriers, 11.7 and 11.3~kJ~mol$^{-1}$, respectively. Their rate
constants are plotted in Figure~\ref{fig:rates}. They are rather similar,
which can be expected from the similar barriers. Overall, the rate constants
of both reactions are rather low: at 100~K they are $8.5\cdot
10^{-16}$~cm$^3$~s$^{-1}$ and $2.4\cdot 10^{-16}$~cm$^3$~s$^{-1}$ while at
50~K, the lowest temperature for which we performed instanton calculations,
they are merely $1.1\cdot 10^{-17}$~cm$^3$~s$^{-1}$ and $3.7\cdot
10^{-18}$~cm$^3$~s$^{-1}$. In sum, our calculations reveal that propynal is rather stable 
against reaction with H atoms, unlike propadienone, which will be efficiently destroyed by atomic hydrogen 
on both grain surfaces and in the gas.

\section{Astrochemical Implications} \label{sec:astrochem}

Atomic hydrogen is a known component of even  dense molecular clouds, where it is produced mainly via the cosmic ray-driven dissociation of \ce{H2} \citep{padovani_production_2018}. Thus produced, these atoms can adsorb onto the surface of interstellar dust grains, where their high mobilities make reactions between them and other grain species one of the dominant formation routes for complex organic molecules in the ISM \citep{herbst_complex_2009}. Cosmic rays can also drive the formation of atomic hydrogen within interstellar ices via radiolytic dissociation \citep{shingledecker_cosmic-ray-driven_2018,shingledecker_general_2018}. Therefore, given the ubiquity of H in the gas, as well as both on and in dust-grain ice mantles, our finding that H atom addition to \ce{CH2CCO} \eqref{r4} is barrierless and exothermic provides a compelling explanation as to why this most stable \ce{H2C3O} isomer has remained undetected in sources over a wide range of physical conditions \citep{loomis_investigating_2015,loison_interstellar_2016}. 

Our hypothesis that reaction \eqref{r4} is indeed the long-sought process underlying the consistent non-detections of propadienone is further supported by both previous observational and experimental studies. For example, in \citet{zhou_pathways_2008} it was shown that, in a mixed CO:\ce{C2H2} ice at 10 K irradiated by high-energy electrons under ultra-high vacuum, the formation of both cyclopropenone and propynal could be observed, though interestingly, no firm detection of propadienone could be made. There, atomic hydrogen was efficiently formed throughout the ice via the dissociation of acetylene, which could then quickly destroy \ce{CH2CCO} even at the very low temperatures at which the experiment was carried out. 

Another finding suggestive of the importance of reaction \eqref{r4} was the detection by \citet{hollis_green_2004} of propenal (\ce{CH2CHCHO}) in Sgr B2(N), where \ce{HCCCHO} and \ce{c-H2C3O} have been seen \citep{loomis_investigating_2015}. In the paper reporting their detection, Hollis \textit{et al.} proposed the following formation route for \ce{CH2CHCHO}:

\begin{equation}
    \ce{HCCCHO + 2H -> CH2CHCHO}.
    \label{r6}
    \tag{R15}
\end{equation}

\noindent
However, as the results of our calculations show, a much more energetically favored precursor is in fact \ce{CH2CCO}. If 
reaction \eqref{r4} followed by \eqref{r5}, as opposed to \eqref{r6} is the dominant formation route for \ce{CH2CHCHO} then, in 
a sense, propadienone might be hiding in plain sight as propenal, rather than missing. 

More generally, our results support the claim made by \citet{loomis_investigating_2015} that molecular abundances in interstellar environments, even in hot cores like Sgr B2(N), are ultimately kinetically controlled, i.e.  governed by reaction barriers rather than the thermodynamic stabilities of individual species. Moreover, as one can glean from an overview of the data in Table~\ref{tab:all}, it is often not possible to intuit the presence or size of such barriers, particularly for neutral-neutral reactions not involving two radicals. Thus, detailed quantum chemical calculations are essential astrochemical tools that can, as shown here, shed light on the underlying processes which give rise to seemingly perplexing observational results.  

Finally, these data both strengthen the chemical connection between unsaturated carbon chain species like \ce{HC3O} and nearly saturated organic molecules, such as \ce{CH2CHCHO}, and suggest that other members of the HC$_n$O ($n=3-7$) family - as well as perhaps similiar carbon-chain species like the ubiquitous cyanopolyynes - might likewise serve as backbones for more complex molecules \citep{mcguire_detection_2018}. 

\section{Conclusions and Outlook} \label{sec:conc}

We have carried out calculations of reactions between atomic hydrogen and \ce{HC3O}, \ce{CH2CCO}, and \ce{HCCCHO}. Our main findings are the following:

\begin{enumerate}
    \item that \ce{H + HC3O}, \eqref{r1}-\eqref{r2}, is both barrierless and exothermic, and leads to either  propadienone (\ce{CH2CCO}) or propynal (\ce{HCCCHO}), with the orientation of the reactants being the main factor influencing which of these two products is formed, 
    \item that the reactivity of propynal and propadienone with H are starkly different, with only \ce{H + CH2CCO -> CH2CHCO} \eqref{r4} being both barrierless and exothermic,
    \item that the above finding serves as a compelling explanation as to why attempts to detect propadienone have been consistently negative and,
    \item that the subsequent barrierless exothermic reaction of \ce{CH2CHCO} with H, shown in \eqref{r5}, yields propenal (\ce{CH2CHCHO}) which has been observed in Sgr B2(N).
\end{enumerate}

These findings are in agreement with recent work by \citet{garrod_exploring_2017}, who examined the formation mechanisms of both the branched (\ce{i-C3H7CN}) and straight-chain (\ce{n-C3H7CN}) forms of propyl cyanide in detail. As with propadienone, the more stable isomer (\ce{i-C3H7CN}) was found to be less abundant than \ce{n-C3H7CN} \citep{belloche_detection_2014}. Interestingly, it was found by \citet{garrod_exploring_2017} that kinetic factors - specifically the rates of reaction with H and CN and the relative efficiencies of the addition of these radicals to either secondary or terminal carbon atoms - were essential for accurately reproducing the $i$:$n$ ratio.

Thus, though the focus of this work was on explaining the consistent non-detections of propadienone and not on disproving the already questionable MEP, our results further reinforce the central role of kinetics in understanding the behavior of interstellar isomers, even in comparatively warm environments like star forming regions. Unfortunately, this makes interpreting observed abundances more challenging since even similar species, like propynal and propadienone, can display quite different reactivities. Thus, as the variety and complexity of known interstellar molecules continues to increase, so too will the importance of detailed experiments or quantum chemical calculations in understanding the chemical basis underlying observational results. 

\acknowledgments

CNS gratefully acknowledges the support of the Alexander von Humboldt
Foundation. This work was financially supported by the European Union's
Horizon 2020 research and innovation program (Grant Agreement Number 646717,
TUNNELCHEM).

\software{Turbomole \citep{TURBOMOLE}, ChemShell \citep{she03,met14}, DL-Find \citep{dlf}, Molpro \citep{molpro2010}}

\bibliography{references}

\begin{thebibliography}{}
\expandafter\ifx\csname natexlab\endcsname\relax\def\natexlab#1{#1}\fi
\providecommand{\url}[1]{\href{#1}{#1}}

\bibitem[{Adler {et~al.}(2007)Adler, Knizia, \& Werner}]{ccf12}
Adler, T.~B., Knizia, G., \& Werner, H.-J. 2007, J. Chem. Phys., 127, 221106

\bibitem[{Belloche {et~al.}(2014)Belloche, Garrod, Müller, \&
  Menten}]{belloche_detection_2014}
Belloche, A., Garrod, R.~T., Müller, H. S.~P., \& Menten, K.~M. 2014, Science,
  345, 1584.
\newblock \url{http://adsabs.harvard.edu/abs/2014Sci...345.1584B}

\bibitem[{Coleman(1977)}]{col77}
Coleman, S. 1977, Phys. Rev. D, 15, 2929

\bibitem[{Cooksy {et~al.}(1995)Cooksy, Tao, Klemperer, \& Thaddeus}]{coo95}
Cooksy, A.~L., Tao, F.-M., Klemperer, W., \& Thaddeus, P. 1995, J. Phys. Chem.,
  99, 11095

\bibitem[{Cordiner {et~al.}(2017)Cordiner, Charnley, Kisiel, McGuire, \&
  Kuan}]{cordiner_deep_2017}
Cordiner, M.~A., Charnley, S.~B., Kisiel, Z., McGuire, B.~A., \& Kuan, Y.-J.
  2017, The Astrophysical Journal, 850, 187.
\newblock \url{http://adsabs.harvard.edu/abs/2017ApJ...850..187C}

\bibitem[{Ekern {et~al.}(1996)Ekern, Szczepanski, \& Vala}]{eke96}
Ekern, S., Szczepanski, J., \& Vala, M. 1996, J. Phys. Chem., 100, 16109

\bibitem[{Garrod {et~al.}(2017)Garrod, Belloche, Müller, \&
  Menten}]{garrod_exploring_2017}
Garrod, R.~T., Belloche, A., Müller, H. S.~P., \& Menten, K.~M. 2017,
  Astronomy and Astrophysics, 601, A48.
\newblock \url{http://adsabs.harvard.edu/abs/2017A%26A...601A..48G}

\bibitem[{Herbst \& van Dishoeck(2009)}]{herbst_complex_2009}
Herbst, E., \& van Dishoeck, E.~F. 2009, Annual Review of Astronomy and
  Astrophysics, 47, 427

\bibitem[{Hollis {et~al.}(2004)Hollis, Jewell, Lovas, Remijan, \&
  Møllendal}]{hollis_green_2004}
Hollis, J.~M., Jewell, P.~R., Lovas, F.~J., Remijan, A., \& Møllendal, H.
  2004, The Astrophysical Journal Letters, 610, L21.
\newblock \url{http://adsabs.harvard.edu/abs/2004ApJ...610L..21H}

\bibitem[{Hollis {et~al.}(2006)Hollis, Remijan, Jewell, \&
  Lovas}]{hollis_cyclopropenone_2006}
Hollis, J.~M., Remijan, A.~J., Jewell, P.~R., \& Lovas, F.~J. 2006, The
  Astrophysical Journal, 642, 933.
\newblock \url{http://adsabs.harvard.edu/abs/2006ApJ...642..933H}

\bibitem[{Irvine {et~al.}(1988)Irvine, Friberg, Hjalmarson, Ishikawa, Kaifu,
  Kawaguchi, Madden, Matthews, Ohishi, Saito, Suzuki, Thaddeus, Turner,
  Yamamoto, \& Ziurys}]{irvine_identification_1988}
Irvine, W.~M., Friberg, P., Hjalmarson, A., {et~al.} 1988, The Astrophysical
  Journal Letters, 334, L107.
\newblock \url{http://adsabs.harvard.edu/abs/1988ApJ...334L.107I}

\bibitem[{Karton \& Talbi(2014)}]{karton_pinning_2014}
Karton, A., \& Talbi, D. 2014, Chemical Physics, 436, 22.
\newblock \url{http://adsabs.harvard.edu/abs/2014CP....436...22K}

\bibitem[{K\"astner {et~al.}(2009)K\"astner, Carr, Keal, Thiel, Wander, \&
  Sherwood}]{dlf}
K\"astner, J., Carr, J.~M., Keal, T.~W., {et~al.} 2009, The Journal of Physical
  Chemistry A, 113, 11856

\bibitem[{Komornicki {et~al.}(1981)Komornicki, Dykstra, Vincent, \&
  Radom}]{kom81}
Komornicki, A., Dykstra, C.~E., Vincent, M.~A., \& Radom, L. 1981, J. Am. Chem.
  Soc., 103, 1652

\bibitem[{Kryvohuz(2013)}]{kry13}
Kryvohuz, M. 2013, J. Chem. Phys., 138, 244114

\bibitem[{Lattelais {et~al.}(2009)Lattelais, Pauzat, Ellinger, \&
  Ceccarelli}]{lattelais_interstellar_2009}
Lattelais, M., Pauzat, F., Ellinger, Y., \& Ceccarelli, C. 2009, The
  Astrophysical Journal Letters, 696, L133.
\newblock \url{http://adsabs.harvard.edu/abs/2009ApJ...696L.133L}

\bibitem[{Lattelais {et~al.}(2010)Lattelais, Pauzat, Ellinger, \&
  Ceccarelli}]{lattelais_new_2010}
---. 2010, Astronomy and Astrophysics, 519, A30.
\newblock \url{http://adsabs.harvard.edu/abs/2010A%26A...519A..30L}

\bibitem[{Lattelais {et~al.}(2011)Lattelais, Bertin, Mokrane, Romanzin,
  Michaut, Jeseck, Fillion, Chaabouni, Congiu, Dulieu, Baouche, Lemaire,
  Pauzat, Pilmé, Minot, \& Ellinger}]{lattelais_differential_2011}
Lattelais, M., Bertin, M., Mokrane, H., {et~al.} 2011, Astronomy and
  Astrophysics, 532, A12.
\newblock \url{http://adsabs.harvard.edu/abs/2011A%26A...532A..12L}

\bibitem[{Loison {et~al.}(2016)Loison, Agúndez, Marcelino, Wakelam, Hickson,
  Cernicharo, Gerin, Roueff, \& Guélin}]{loison_interstellar_2016}
Loison, J.-C., Agúndez, M., Marcelino, N., {et~al.} 2016, Monthly Notices of
  the Royal Astronomical Society, 456, 4101.
\newblock \url{http://adsabs.harvard.edu/abs/2016MNRAS.456.4101L}

\bibitem[{Loomis {et~al.}(2015)Loomis, McGuire, Shingledecker, Johnson, Blair,
  {Amy Robertson}, \& Remijan}]{loomis_investigating_2015}
Loomis, R.~A., McGuire, B.~A., Shingledecker, C., {et~al.} 2015, The
  Astrophysical Journal, 799, 34.
\newblock \url{http://stacks.iop.org/0004-637X/799/i=1/a=34}

\bibitem[{Maclagan {et~al.}(1995)Maclagan, McEwan, \& Scott}]{mac95}
Maclagan, R.~G., McEwan, M.~J., \& Scott, G. 1995, Chem. Phys. Lett., 240, 185
  .
\newblock
  \url{http://www.sciencedirect.com/science/article/pii/000926149500503V}

\bibitem[{McConnell \& K\"astner(2017)}]{mcc17a}
McConnell, S., \& K\"astner, J. 2017, J. Comput. Chem., 38, 2570.
\newblock \url{http://dx.doi.org/10.1002/jcc.24914}

\bibitem[{{McGuire}(2018)}]{mcguire_census_2018}
{McGuire}, B.~A. 2018, \apjs, 239, 17

\bibitem[{McGuire {et~al.}(2018)McGuire, Burkhardt, Kalenskii, Shingledecker,
  Remijan, Herbst, \& McCarthy}]{mcguire_detection_2018}
McGuire, B.~A., Burkhardt, A.~M., Kalenskii, S., {et~al.} 2018, Science, 359,
  202.
\newblock \url{http://science.sciencemag.org/content/359/6372/202}

\bibitem[{McGuire {et~al.}(2017)McGuire, Burkhardt, Shingledecker, Kalenskii,
  {Eric Herbst}, Remijan, \& McCarthy}]{mcguire_detection_2017}
McGuire, B.~A., Burkhardt, A.~M., Shingledecker, C.~N., {et~al.} 2017, The
  Astrophysical Journal Letters, 843, L28.
\newblock \url{http://stacks.iop.org/2041-8205/843/i=2/a=L28}

\bibitem[{Meisner {et~al.}(2017)Meisner, Lamberts, \&
  Kästner}]{meisner_atom_2017}
Meisner, J., Lamberts, T., \& Kästner, J. 2017, ACS Earth and Space Chemistry,
  1, 399.
\newblock \url{http://dx.doi.org/10.1021/acsearthspacechem.7b00052}

\bibitem[{Metz {et~al.}(2014)Metz, K\"astner, Sokol, Keal, \& Sherwood}]{met14}
Metz, S., K\"astner, J., Sokol, A.~A., Keal, T.~W., \& Sherwood, P. 2014, WIREs
  Comput. Mol. Sci., 4, 101

\bibitem[{Miller(1975)}]{mil75}
Miller, W.~H. 1975, J. Chem. Phys., 62, 1899

\bibitem[{{Padovani} {et~al.}(2018){Padovani}, {Galli}, {Ivlev}, {Caselli}, \&
  {Ferrara}}]{padovani_production_2018}
{Padovani}, M., {Galli}, D., {Ivlev}, A.~V., {Caselli}, P., \& {Ferrara}, A.
  2018, \aap, 619, A144

\bibitem[{Peterson {et~al.}(2008)Peterson, Adler, \& Werner}]{pet08}
Peterson, K.~A., Adler, T.~B., \& Werner, H.-J. 2008, J. Chem. Phys., 128,
  084102

\bibitem[{Rommel {et~al.}(2011)Rommel, Goumans, \& K\"astner}]{rom11}
Rommel, J.~B., Goumans, T. P.~M., \& K\"astner, J. 2011, J. Chem. Theory
  Comput., 7, 690

\bibitem[{Sch{\"a}fer {et~al.}(1994)Sch{\"a}fer, Huber, \& Ahlrichs}]{tzvp}
Sch{\"a}fer, A., Huber, C., \& Ahlrichs, R. 1994, The Journal of Chemical
  Physics, 100, 5829

\bibitem[{Scott \& Radom(2000)}]{sco00}
Scott, A., \& Radom, L. 2000, J. Mol. Struct., 556, 253 .
\newblock
  \url{http://www.sciencedirect.com/science/article/pii/S0022286000006402}

\bibitem[{Sherwood {et~al.}(2003)Sherwood, de~Vries, Guest, Schreckenbach,
  Catlow, French, Sokol, Bromley, Thiel, Turner, Billeter, Terstegen, Thiel,
  Kendrick, Rogers, Casci, Watson, King, Karlsen, Sj{\o}voll, Fahmi,
  Sch{\"a}fer, \& Lennartz}]{she03}
Sherwood, P., de~Vries, A.~H., Guest, M.~F., {et~al.} 2003, J. Mol. Struct.
  (THEOCHEM), 632, 1

\bibitem[{Shingledecker \& Herbst(2018)}]{shingledecker_general_2018}
Shingledecker, C.~N., \& Herbst, E. 2018, Physical Chemistry Chemical Physics,
  20, 5359.
\newblock
  \url{http://pubs.rsc.org/en/content/articlelanding/2018/cp/c7cp05901a}

\bibitem[{Shingledecker {et~al.}(2018)Shingledecker, Tennis, Gal, \&
  Herbst}]{shingledecker_cosmic-ray-driven_2018}
Shingledecker, C.~N., Tennis, J., Gal, R.~L., \& Herbst, E. 2018, The
  Astrophysical Journal, 861, 20.
\newblock \url{http://stacks.iop.org/0004-637X/861/i=1/a=20}

\bibitem[{TURBOMOLE(2018)}]{TURBOMOLE}
TURBOMOLE. 2018, {V7.1}, a development of {University of Karlsruhe} and
  {Forschungszentrum Karlsruhe GmbH}, 1989-2007, {TURBOMOLE GmbH}, since 2007;
  available from \\ {\tt http://www.turbomole.com}., ,

\bibitem[{Wang \& Cooksy(1996)}]{radref}
Wang, H., \& Cooksy, A. 1996, Chem. Phys., 213, 139

\bibitem[{Werner {et~al.}(2010)Werner, Knowles, Manby, {Sch\"{u}tz}, Celani,
  Knizia, Korona, Lindh, Mitrushenkov, Rauhut, Adler, Amos, Bernhardsson,
  Berning, Cooper, Deegan, Dobbyn, Eckert, Goll, Hampel, Hesselmann, Hetzer,
  Hrenar, Jansen, K\"oppl, Liu, Lloyd, Mata, May, McNicholas, Meyer, Mura,
  Nicklass, Palmieri, Pfl\"uger, Pitzer, Reiher, Shiozaki, Stoll, Stone,
  Tarroni, Thorsteinsson, Wang, \& Wolf}]{molpro2010}
Werner, H.-J., Knowles, P.~J., Manby, F.~R., {et~al.} 2010, MOLPRO, version
  2010.2, a package of ab initio programs, , , see www.molpro.net

\bibitem[{Zhao \& Truhlar(2005)}]{pw6}
Zhao, Y., \& Truhlar, D.~G. 2005, The Journal of Physical Chemistry A, 109,
  5656

\bibitem[{Zhou {et~al.}(2008)Zhou, Kaiser, Gao, Chang, Liang, \&
  Yung}]{zhou_pathways_2008}
Zhou, L., Kaiser, R.~I., Gao, L.~G., {et~al.} 2008, The Astrophysical Journal,
  686, 1493

\end{thebibliography}

\end{document}